\documentclass[twocolumn]{aastex61}

\usepackage{amsmath}
\usepackage{amsfonts}
\usepackage{graphicx}
\usepackage{natbib}
\usepackage{txfonts}
\usepackage{color}
\usepackage{wrapfig}
\usepackage{amssymb}
\usepackage{color}

\shorttitle{Adsorption energies of C, N, and O on water ice}
\shortauthors{Shimonishi and Nakatani et al.}

\begin{document}

\title{Adsorption energies of carbon, nitrogen, and oxygen atoms on the low-temperature amorphous water ice: A systematic estimation from quantum chemistry calculations}

\correspondingauthor{Naoki Nakatani and Takashi Shimonishi} 
\email{naokin@tmu.ac.jp} 
\email{shimonishi@astr.tohoku.ac.jp}

\author{Takashi Shimonishi} 
\altaffiliation{These authors contributed equally to this work.} 
\affiliation{Frontier Research Institute for Interdisciplinary Sciences, Tohoku University, Aramakiazaaoba 6-3, Aoba-ku, Sendai, Miyagi, 980-8578, Japan} 
\affiliation{Astronomical Institute, Tohoku University, Aramakiazaaoba 6-3, Aoba-ku, Sendai, Miyagi, 980-8578, Japan} 

\author{Naoki Nakatani} 
\altaffiliation{These authors contributed equally to this work.} 
\affiliation{Institute for Catalysis, Hokkaido University, N21W10 Kita-ku, Sapporo, Hokkaido 001-0021, Japan} 
\affiliation{Department of Chemistry, Graduate School of Science and Engineering, Tokyo Metropolitan University, 1-1 Minami-Osawa, Hachioji, Tokyo 192-0397, Japan} 

\author{Kenji Furuya} 
\affiliation{Center for Computational Sciences, The University of Tsukuba, 1-1-1, Tennodai, Tsukuba, Ibaraki 305-8577, Japan} 

\author{Tetsuya Hama} 
\affiliation{Institute for Low Temperature Science, Hokkaido University, N19W8 Kita-ku, Sapporo, Hokkaido 060-0819, Japan}

%% Note that the \and command from previous versions of AASTeX is now
%% depreciated in this version as it is no longer necessary. AASTeX 
%% automatically takes care of all commas and "and"s between authors names.

%% AASTeX 6.1 has the new \collaboration and \nocollaboration commands to
%% provide the collaboration status of a group of authors. These commands 
%% can be used either before or after the list of corresponding authors. The
%% argument for \collaboration is the collaboration identifier. Authors are
%% encouraged to surround collaboration identifiers with ()s. The 
%% \nocollaboration command takes no argument and exists to indicate that
%% the nearby authors are not part of surrounding collaborations.

%% Mark off the abstract in the ``abstract'' environment. 
\begin{abstract}
We propose a new simple computational model to estimate adsorption energies of atoms and molecules to low-temperature amorphous water ice, and we present the adsorption energies of carbon ($^3P$), nitrogen ($^4S$), and oxygen ($^3P$) atoms based on quantum chemistry calculations.
The adsorption energies were estimated to be 14100 $\pm$ 420 K for carbon, 400 $\pm$ 30 K for nitrogen, and 1440 $\pm$ 160 K for oxygen. 
The adsorption energy of oxygen is well consistent with experimentally reported value.
We found that the binding of a nitrogen atom is purely physisorption, while that of a carbon atom is chemisorption in which a chemical bond to an O atom of a water molecule is formed.
That of an oxygen atom has a dual character both physisorption and chemisorption.
The chemisorption of atomic carbon also implies a possibility of further chemical reactions to produce molecules bearing a C--O bond, 
while it may hinder the formation of methane on water ice via sequential hydrogenation of carbon atoms. 
These would be of a large impact to the chemical evolution of carbon species in interstellar environments.
We also investigated effects of the newly calculated adsorption energies onto chemical compositions of cold dense molecular clouds with the aid of gas-ice astrochemical simulations. 
We found that abundances of major nitrogen-bearing molecules, such as N$_2$ and NH$_3$, are significantly altered by applying the calculated adsorption energy, because nitrogen atoms can thermally diffuse on surfaces even at 10 K. 
\end{abstract}

%% Keywords should appear after the \end{abstract} command. 
%% See the online documentation for the full list of available subject
%% keywords and the rules for their use.

%% From the front matter, we move on to the body of the paper.
%% Sections are demarcated by \section and \subsection, respectively.
%% Observe the use of the LaTeX \label
%% command after the \subsection to give a symbolic KEY to the
%% subsection for cross-referencing in a \ref command.
%% You can use LaTeX's \ref and \label commands to keep track of
%% cross-references to sections, equations, tables, and figures.
%% That way, if you change the order of any elements, LaTeX will
%% automatically renumber them.

%% We recommend that authors also use the natbib \citep
%% and \citet commands to identify citations.  The citations are
%% tied to the reference list via symbolic KEYs. The KEY corresponds
%% to the KEY in the \bibitem in the reference list below. 

\section{Introduction}
Formation of molecules on grain surfaces plays an essential role in the chemical evolution of cold and dense molecular clouds. 
Adsorption energy ($E_\mathrm{ads.}$) and diffusion activation energy ($E_\mathrm{diff.}$) of surface atoms and molecules are one of the important parameters that control the efficiency of grain surface reactions. 
Astrochemical simulations of gas-grain chemistry suggest that chemical compositions of dense molecular clouds are highly dependent on the $E_\mathrm{ads.}$ assumed in the simulation \citep{Wak17,Pen17}. 
$E_\mathrm{diff.}$ can be estimated as a fraction of $E_\mathrm{ads.}$ of surfaces species; the ratio of $E_\mathrm{diff.}$ to $E_\mathrm{ads.}$ often ranges from 0.3 to 1.0 \citep[e.g.,][]{Sla74,Med11,Kar14,Cup17}. 
Accurate information on adsorption energies of major surface species is thus crucial for astrochemical modeling of dense molecular cloud chemistry. 

The values of $E_\mathrm{ads.}$ of stable molecules have been experimentally determined using thermal desorption spectroscopy such as temperature-programmed desorption (TPD) methods \citep[e.g.,][]{Bur10,Ham13}. 
However, TPD methods are not appropriate for reactive atoms (e.g., H), because they can barrierlessly recombine to form stable molecules on surface before thermal desorption (H + H $\to$ H$_2$). 
In addition, direct detection of atoms is difficult using quadrupole mass spectrometer with electron ionization. 
$E_\mathrm{ads.}$ and $E_\mathrm{diff.}$ of atoms have been indirectly obtained from the analysis of TPD spectra of molecular products using a rate-equation model with $E_\mathrm{ads.}$ and $E_\mathrm{diff.}$ as parameters. 
However, the experimental results can be contradictory, depending on the difference in the experimental conditions (e.g., the incident flux of the atoms) and assumptions about the surface coverage of atoms \citep{Man01,Hor03,Pir04,Per05,Mat08,Vid06,Ham13}. 

To overcome this problem, the photo-stimulated desorption and resonance-enhanced multiphoton ionization (PSD-REMPI) method was developed to directly investigate adsorption and diffusion of H atoms on water ice \citep{Wat10,Ham12,Kuw15}, but the PSD-REMPI method is yet to be applied to other atoms such as C, N, and O. 
Recent laboratory studies have suggested that the $E_\mathrm{ads.}$ of atomic oxygen (O) on interstellar dust analogues (1400--1700 K) significantly deviates  from the traditionally adopted value of 800 K estimated by \citet{Tie82} on the basis of its polarizability \citep{War12,Kim14,He15,Min16}. 
An experimental approach is also applied to the adsorption energy of atomic nitrogen in \citet{Min16}, while laboratory measurement for atomic carbon is not reported so far. 
These experimental studies suggest an importance to revisiting $E_\mathrm{ads.}$ of surface species with the aid of the latest computational techniques.

Theoretically estimating the adsorption energies of atoms and molecules on water ice is a still challenging work even though we are able to use high-performance computers, because interstellar phenomena takes at least thousands of years to complete, which are impossible to implement into the current computer resources.
Therefore, a suitable computational modeling with reliable approximations is indispensable in practical simulations for interstellar chemistry.

\citet{Buc91}, \citet{AlH02}, and \citet{AlH07} reported molecular dynamics simulation to theoretically investigate a collision process of a hydrogen atom to crystalline and amorphous water ice. 
They employed potential parameters which were generated from ab-initio quantum chemical calculation. 
Many subsequent simulations extensively studied sticking probability, adsorption, and diffusion of H atoms on water ice \citep[e.g.,][]{Vee14,Dup16,Asg17,Sen17}. 
The mobility of an O atom in amorphous ice at low temperatures was also recently studied using classical molecular dynamics \citep{MLee14}. 
However, classical molecular dynamics cannot capture chemical reaction since bond reformation is not involved in the computational models. 
Theoretical approaches based on quantum chemical calculations are thus highly desirable for understanding the adsorption of atoms, especially C, N, and O atoms, on water ice.

Quantum chemical calculations are now often used to theoretically investigate chemical reactions. 
For example, \citet{Ozk12} reported a theoretical work on the chemical evolution of a carbon atom with a water molecule based on ab-initio quantum chemical calculations. 
They considered chemical reactions between singlet and triplet carbon atoms and one water molecule to formation of a formaldehyde. 
Though they employed high-level quantum chemical theories to explore the chemical reactions, their computational model involved only one water molecule, which was far from the realistic condition in interstellar ices possibly due to computational limitations. 

Recently, \citet{Wak17} also reported binding energies of various atoms and molecules to one water molecule based on quantum chemical calculations. 
They compared the calculated binding energies with experimental values for some molecules and concluded that the calculated binding energies to one water molecule are proportional to the experimental values. 
However, it had been also mentioned that this is only true if a covalent bond does not form between adsorbed species and the water molecule. 

Consequently, estimation of binding energies of bare atomic radicals such as carbon, nitrogen, oxygen, and so on, is not simply doable by interpolating the one water model to the experiments. 
In this work, we wish to propose a new calculation model to incorporate both statistical and quantum effects in the estimation of the binding energies of atoms and molecules to the amorphous solid water (ASW) surface, and to demonstrate our model to compute the binding energies of carbon ($^3P$), nitrogen ($^4S$), and oxygen ($^3P$). 

In this paper, we first describe computational model which is applicable to estimate binding energies of atoms and molecules in interstellar chemistry (Section \ref{sec_comp}). 
Next, we demonstrate our model to compute binding energies of C, N, and O atoms on the ASW (Section \ref{sec_Eb}). 
With these binding energies, we perform simulations based on the rate-equation method (Section \ref{sec_Method_RE}) and discuss the impact on chemical compositions of dense molecular clouds (Section \ref{sec_Results_RE}). 
Finally, we close our discussions with concluding remarks and future perspectives (Section \ref{sec_concl}).

\section{Computational details} \label{sec_comp} 
\begin{figure*}[!]
\begin{center}
\includegraphics[width=15cm]{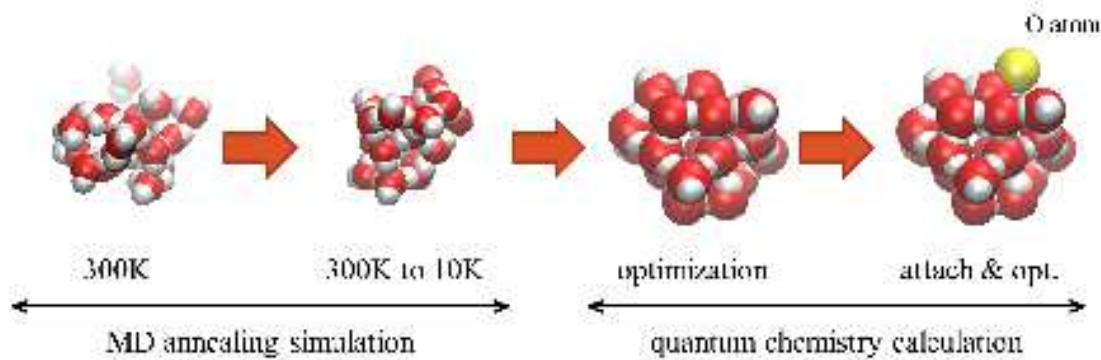}
\caption{Computational steps to construct an amorphous ice cluster model and its atom-adsorbed structure.}
\label{fig_h2o_clust}
\end{center}
\end{figure*}

We present a new computational model of ASW to compute absorption energy, which is simple but able to take into account statistical features in interstellar environments. 
A key hypothesis is that an adsorbent will occupy the most stable site in a local region of ASW surface during the very-long period of time, although motions of atoms and molecules are extremely slow because of the low-temperature condition in interstellar medium. 
In fact, the TPD experiments show that physisorbed species (at least for nonpolar molecules such as H$_2$, D$_2$, and N$_2$) are bound to deep potential sites on the ASW surface following diffusion prior to thermal desorption \citep{Kim01,Hor05,Ami06,Ami07,Fil09}. 
To capture these features in the adsorption energy calculation, we took three steps to construct our computational model: 
(1) several ASW clusters are generated from MD-annealing calculations, 
(2) the adsorbent is randomly added to each ASW cluster and optimize the geometry using quantum chemistry calculations, and
(3) the largest adsorption energies for each cluster are averaged, as shown schematically in Figure \ref{fig_h2o_clust}. 
All the MD-calculations were performed using Amber 14 program package \citep{amber14} and all the quantum chemistry calculations were performed using Gaussian 09 program package \citep{gaussian}. 
In the following subsections, we would like to explain details of each step.
%
%%%%%%%%%%
%%%%%%%%%%

\subsection{MD annealing simulation to construct amorphous solid water (ASW) clusters}
First, we carried out MD-annealing calculations using classical force-fields to perform a water cluster as a model of the ASW surface.
We considered 20 water molecules with the TIP3P model and performed droplet simulation with the spherical constraint of 20 $\AA$ to prevent escape of water molecules from the cluster.
After 100 ps simulation at 300 K to achieve equilibrium, we took 11 different structures each of which was annealed to 10 K as initial guesses of quantum chemistry calculations.
Next, these 11 different structures of water clusters were fully optimized on the basis of density functional theory (DFT) calculations using $\omega$-B97XD functional \citep{chai2008}, in which the van der Waals (vdW) interaction is empirically incorporated.
A valence triple-$\zeta$ plus polarization and diffuse functions, namely 6-311+G(d, p) basis sets, were adopted for both the geometry optimization and the adsorption energy evaluation.
From these optimized structures, we discarded 2 structures because the annealing process failed to form a cluster.
Consequently, we considered 9 different structures of water clusters.

Due to fluctuation of water molecules in a long-time period, our scheme assumes that the ASW achieves a sort of equilibrium condition even though it is at the extremely low-temperature.
Sampling several cluster geometries captures different region of the real ASW surface, and therefore, averaging the adsorption energies over clusters will be a good approximation to the average adsorption energy in the real ASW surface.

\subsection{Adsorption energy evaluation}
For 9 selected structures of [H$_2$O]$_{20}$, we randomly added C, N, or O atom around the surface area of the cluster and fully optimized the geometry using the same DFT functional and the same basis sets.
We considered 10 different ``trials'' for each 9 different ``samples'' of [H$_2$O]$_{20}$, and therefore, 90 different structures of the adsorbed clusters were examined.
Finally, we chose the largest adsorption energies from each 10 different trials and averaged them over 9 different samples to compute the adsorption energy.
The largest adsorption energy from 10 different trials can capture the locally stable site in which the adsorbent stays very long-time period.

\section{Calculated adsorption energies of C, N, and O atoms to ASW} \label{sec_Eb} 
Calculated adsorption energies of carbon (3P), nitrogen (4S), and oxygen (3P) atoms to [H$_2$O]$_{20}$ cluster were summarized in Table \ref{tab_ads}, which were compared with those from the previous experimental and computational results.
Our results are qualitatively consistent with adsorption energies estimated in \citet{Wak17} by using DFT/M06-2X calculations for the interaction of the species with one water molecule. 
In the following subsections, we discuss characteristic features of the adsorption for each atom according to calculated adsorption energy, geometry, and electronic structure.
%
%%%%%%%%%%
\begin{deluxetable}{ l c c c }
\tablecaption{Calculated adsorption energy ($E_\mathrm{ads.}$) of C, N, and O atoms on [H$_2$O]$_{20}$ cluster. \label{tab_ads}}
\tabletypesize{\small} 
\tablehead{
\colhead{}   & \colhead{C($^3P$)} & \colhead{N($^4S$)} & \colhead{O($^3P$)}
}
\startdata
$E_\text{ads.}$ & 14100        & 400  & 1440 \\
Std. error          &   420          &  30   &  160 \\
Exptl.                &   N/A          & 720  & 1410 \\
W17                 &  10000       & 1200 & 1700--2200  \\
\enddata
\tablecomments{
Adsorption energies and standard errors are in units of Kelvin. 
Zero-point energy correction was incorporated. 
Experimentally observed adsorption energies of N and O atoms are shown in the third row \citep{Min16}. 
Adsorption energies estimated in \citet{Wak17} by using DFT/M06-2X calculations are shown in the fourth row. 
}
\end{deluxetable}
%%%%%%%%%%

\subsection{Carbon atom} \label{sec_Eb_C}
\begin{figure*}[tbp]
\begin{center}
\includegraphics[width=14cm]{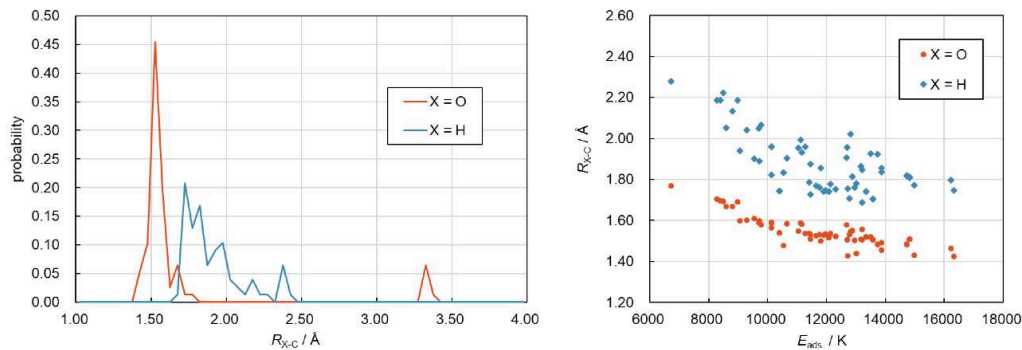}
\caption{Left panel: Histograms of distance between the adsorbed C atom and the nearest-neighbor atom ($R_\text{X-C}$, X = H and O). Converged 77 samples were considered, for every 0.05 $\AA$, and the frequency is divided by number of samples. Right panel: Correlation between the distance ($R_\text{X-C}$) and the adsorption energy of C atom ($E_\text{ads.}$).}
\label{fig_C}
\end{center}
\end{figure*}

The most interesting finding in this work is that the adsorption of a C atom to ASW is assigned to be a chemisorption; the adsorption energy was estimated to be 14100 K (117 kJ/mol) which is apparently larger than that for common physisorption.
In the chemisorption, the carbon atom forms a chemical bond with oxygen atom of water. 
Figure \ref{fig_C} shows that histograms of the distances between the adsorbed C atom and the nearest-neighbor O and H atoms for 77 converged samples ($R_\text{O-C}$ and $R_\text{H-C}$, respectively).
Note that 13 samples were excluded since geometry optimizations were not converged in 12 samples and chemical conversion occurs in the last 1 sample. 
Interestingly, 80$\%$ of samples took the $R_\text{O-C}$ being less than 1.60 $\AA$ which is only a little longer than that 1.43 $\AA$ of known aliphatic C-O bonds.
This is another evidence that the adsorption of the C atom on the ASW surface is assigned to be a chemisorption. 
Furthermore, a somewhat broad peak for the $R_\text{H-C}$ distribution is found in the range of 1.70-2.00 $\AA$. 
Though this is apparently longer than that 1.08 $\AA$ of the typical C-H chemical bond length, this is considered as a typical hydrogen bonding distance. 
Indeed, the O-H bond in which the nearest H atom to the adsorbed C atom is involved directed to the C atom. 
As a result, the adsorbed C atom is incorporated to the hydrogen bond network of the [H$_2$O]$_{20}$ cluster. 
Therefore, both the chemical interaction between the C and the O atoms and the hydrogen bonding interaction between the C and the H atoms contribute the large adsorption energy of the C atom. 
These are also shown by the correlation between the adsorption energy and either the $R_\text{O-C}$ or the $R_\text{H-C}$; the shorter $R_\text{O-C}$ or the shorter $R_\text{H-C}$ gives the larger adsorption energy (Figure \ref{fig_C}). 

The reaction of atomic carbon with water is experimentally suggested for gas-phase reactions in previous studies \citep[e.g.,][]{Ahm83,Hic16}. 
The formation of water-carbon adducts is also suggested for carbon atoms and water molecules in liquid helium droplets at ultra-low-temperature \citep[$\sim$0.4 K,][]{Kra14}. 
Previous quantum chemistry calculations support the observed reactivity of atomic carbon with water \citep{Ahm83,Ozk12}. 
\citet{Hic16} argues the importance of tunneling effects on the enhancement of the C + H$_2$O reaction at low temperature. 
On the other hand, there are studies which report the non-reaction of carbon atoms with water in low-temperature ($\sim$10 K) argon matrices \citep{Ort90,Sch06}. 

An impact of our results is that it is strongly suggested computationally that ``chemical reaction occurs'' between the C atom and a water molecule on the ASW surface upon adsorption, and it is of considerable importance in chemical evolution of C-O species in the interstellar space. 

%%%%%%%%%%
%%%%%%%%%%
%
%%%%%%%%%%
\begin{figure*}[tbp]
\begin{center}
\includegraphics[width=14cm]{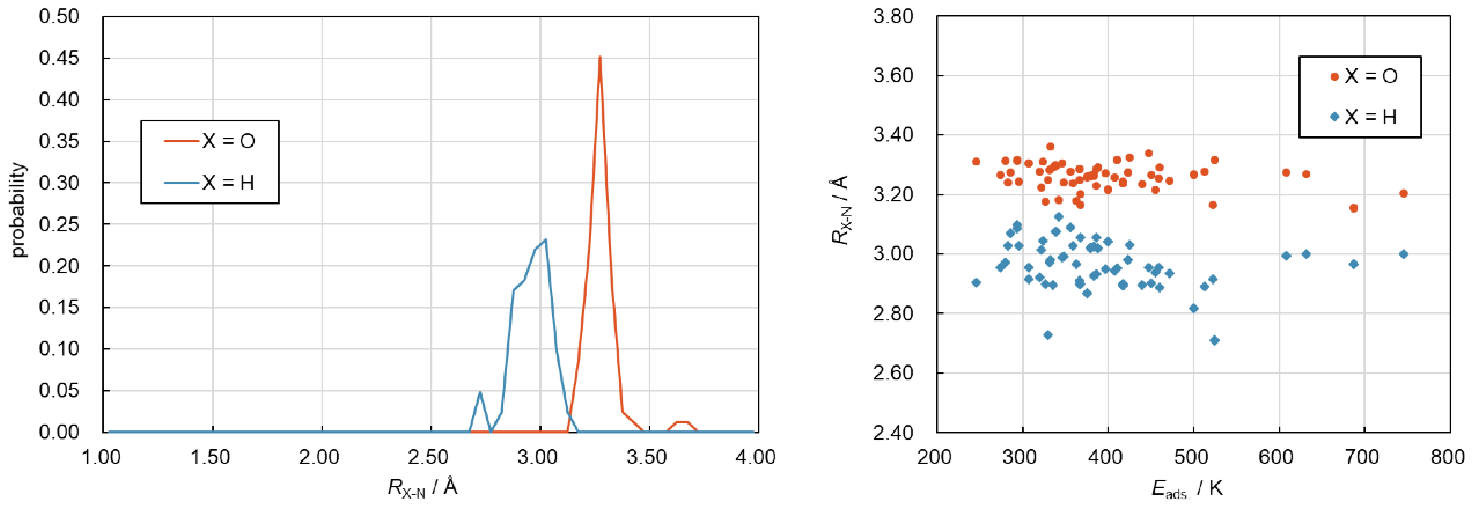}
\caption{Left panel: Histograms of distance between the adsorbed N atom and the nearest-neighbor atom ($R_\text{X-N}$, X = H and O). Converged 82 samples were considered, for every 0.05 $\AA$, and the frequency is divided by number of samples. Right panel: Correlation between the distance ($R_\text{X-N}$) and the adsorption energy of N atom ($E_\text{ads.}$).}
\label{fig_N}
\end{center}
\end{figure*}
%%%%%%%%%%
%
%%%%%%%%%%
\begin{figure*}[tp]
\begin{center}
\includegraphics[width=14cm]{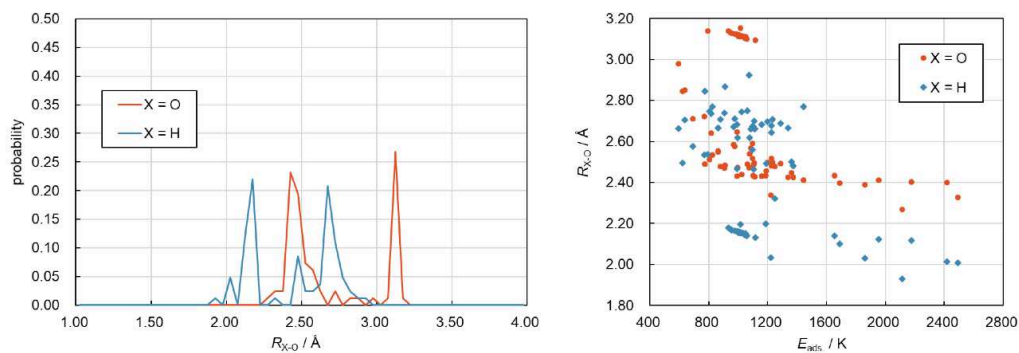}
\caption{Left panel: Histograms of distance between the adsorbed O atom and the nearest-neighbor atom ($R_\text{X-O}$, X = H and O). Converged 82 samples were considered, for every 0.05 $\AA$, and the frequency is divided by number of samples. Right panel: Correlation between the distance ($R_\text{X-O}$) and the adsorption energy of O atom ($E_\text{ads.}$).}
\label{fig_O_1}
\end{center}
\end{figure*}
%%%%%%%%%%

\subsection{Nitrogen atom} \label{sec_Eb_N}
The adsorption energy of an N atom to [H$_2$O]$_{20}$ cluster was estimated to be 400 K (3.33 kJ/mol) which is clearly assigned to be a physisorption. 
The calculated adsorption energy of the N atom somewhat underestimated the experimental value (720 K). 
This underestimation would come from insufficient descriptions of the van der Waals interaction, as we have checked the adsorption energy of a N atom with coupled-cluster calculation (CCSD(T) level of theory, which is known as a gold standard quantum chemistry method for common molecules) in a small model (see the Appendix). 
Consequently, we concluded that the DFT-computed adsorption energy is qualitatively correct at least.
Figure \ref{fig_N} shows that histograms of the distances between the adsorbed N atom and nearest-neighbor O and H atoms for 82 converged samples. 
Note that 8 samples were excluded since geometry optimizations were not converged. 
From the histogram, a peak distribution of the N-O distance appears at $R_\text{O-N}$ = 3.20-3.30 $\AA$ and that of the N-H distance appears at $R_\text{H-N}$ = 2.85-3.00 $\AA$.
These trends can be explained in terms of van der Waals (vdW) radii; estimated N-O and N-H distances are 3.07 $\AA$ and 2.75 $\AA$, respectively (note that the vdW radii of 1.55 $\AA$ for N, 1.52 $\AA$ for O, and 1.20 $\AA$ for H are employed). 
Therefore, we concluded that the N atom adsorbs purely by the vdW interactions between the O and the H atoms of the water cluster. 

Interestingly, neither the N-O distance nor the N-H distance correlates with the adsorption energy (Figure \ref{fig_N}).
This is because the adsorption energy mainly depends on number of coordinate atoms around the adsorbed N atom.
%
%%%%%%%%%%
\begin{figure}[tp]
\begin{center}
\includegraphics[width=7.5cm]{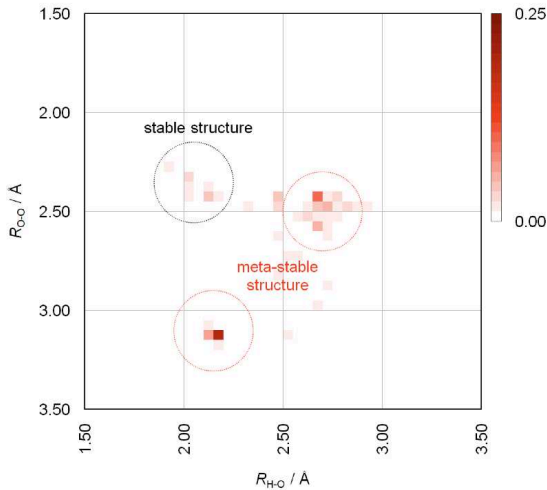}
\caption{Correlation between the $R_\text{O-O}$ and the $R_\text{O-H}$ of the adsorbed O atom on [H$_2$O]$_{20}$.}
\label{fig_O_2}
\end{center}
\end{figure}
%%%%%%%%%%

\subsection{Oxygen atom} \label{sec_Eb_O}
The adsorption energy of an O atom to [H$_2$O]$_{20}$ cluster was computed to be 1440 K (12.0 kJ/mol), which is amazingly close to the experimentally reported value of 1410 K \citep{Min16}. 
Consequently, our results systematically reproduced the adsorption energies of N and O atoms, at least qualitatively, and it is highly expected that the predicted adsorption energy of C atom is to be reliable, too.

From the histogram of distances between the adsorbed O atom and the nearest-neighbor O and H atoms of [H$_2$O]$_{20}$ cluster (Figure \ref{fig_O_1}, left), there are two peaks for each $R_\text{O-O}$ or $R_\text{H-O}$ distribution. 
In the $R_\text{O-O}$ distribution, two peaks are found in the range of 2.40-2.50 $\AA$ and at 3.10 $\AA$. 
In the $R_\text{H-O}$ distribution, two peaks are found in the range of 2.10-2.20 $\AA$ and in the range of 2.65-2.75 $\AA$. 
According to vdW radii, the O-O and the H-O distances are estimated to be 3.04 $\AA$ and 2.72 $\AA$, respectively, and these agree well with the second peaks of the $R_\text{O-O}$ and the $R_\text{H-O}$ distribution. 
On the other hand, correlation between the distance ($R_\text{O-O}$ or $R_\text{H-O}$) and the adsorption energy shows complicated character (Figure \ref{fig_O_1}, right) so that we are unable to understand the adsorbed structure with its stability. 
To make this clear, the correlation between the $R_\text{O-O}$ and the $R_\text{H-O}$ is investigated as shown in Figure \ref{fig_O_2}.
From these results, we found that there are three characteristic structures;
(1) having the short $R_\text{O-O}$ and the long $R_\text{H-O}$,
(2) having the long $R_\text{O-O}$ and the short $R_\text{H-O}$, and
(3) having the short $R_\text{O-O}$ and the short RH-H.
The long $R_\text{O-O}$ or the long $R_\text{H-O}$ distance is understood as that there is only the vdW interaction between the adsorbed O atom and the H$_2$O.
The short $R_\text{O-O}$ or the short $R_\text{H-O}$ distance implies that there is some sort of chemical bonding interaction including the hydrogen bonding interaction, as we seen in the adsorbed C atom although they are very weak rather than the C-O bonding interaction.
Though the structures (1) and (2) are associated with large probability, they are assigned to be meta-stable structures.
Although the structure (3) rarely occurs, this is stable in energy and in our estimation scheme, it gives large contribution to the adsorption energy of the O atom.

\section{Astrochemical implication} \label{sec_RE} 
\subsection{Simulation of dense cloud chemistry with the rate equation method} \label{sec_Method_RE} 
We carried out gas-ice astrochemical simulations to examine the effect of the calculated adsorption energies on chemical compositions of dense molecular clouds. 
We employed a pseudo-time-dependent gas-ice chemistry model, adopting the modified rate equation method \citep{Has92,Gar08a}. 
The chemistry is described by a three-phase model \citep[the gas phase, an icy grain surface, and the bulk ice mantle;][]{Has93}. 
Our chemical network is originally based on that of \citet{Gar06}, in which gas phase reactions, interaction between gas and (icy) grain surface, and surface reactions are included. 
More details can be found in \citet{Fur16,Fur17}.

Simulations are performed for a static dense molecular cloud with the hydrogen nuclei density of $n_\mathrm{H}$ = 2 $\times$ 10$^{5}$ cm$^{-3}$ and the visual extinction of $A_{V}$ = 10 mag. 
Temperatures of both gas and dust are fixed at 10 K or 15 K. 
A standard grain size of 0.1 $\mu$m in radius is assumed, with $\sim$10$^6$ surface binding sites per grain and with the dust-to-gas mass ratio of 0.01. 
Initial gas-phase abundances are shown in Table \ref{tab_initab} for important species; the initial species are assumed to be atoms or atomic ions except for H$_2$ and CO. 
Almost all hydrogen is assumed to be in H$_2$, while half of carbon is assumed to be in CO. 

Two sets of adsorption energies are investigated; Model 1 employs commonly-used adsorption energies \citep[C: 800 K, N: 800 K, O: 1600 K, e.g.,][]{HW13}, whereas Model 2 employs the adsorption energies that are calculated in this work (C: 14100 K, N: 400 K, O: 1440 K, as in Table 1). 
The adsorption energy of atomic hydrogen is set to 350 K in both models. 
It should be noted that our astrochemical models do not consider the formation of a C--O bond upon adsorption of atomic carbon onto water ice for simplicity. 
The ratio of a diffusion energy relative to an adsorption energy is fixed to 0.6. 
The cosmic-ray ionization rate is set to 5.0 $\times$ 10$^{-17}$ s$^{-1}$ \citep{Dal06}. 

\begin{figure}[tbp]
\begin{center}
\includegraphics[width=7.5cm]{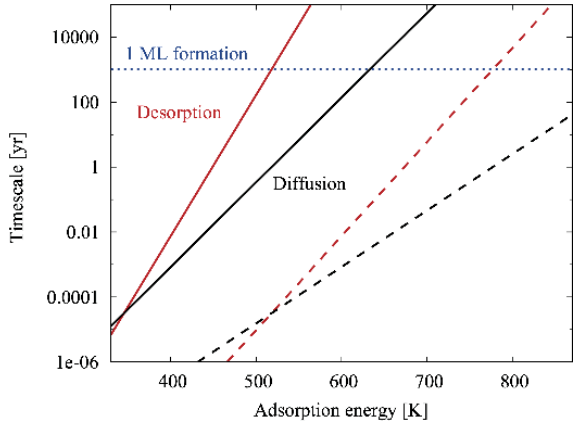}
\caption{Timescales of the desorption (red), diffusion over 10$^6$ sites (black), and formation of a monolayer ice (blue dotted) as a function of adsorption energies. 
The solid lines represent the case of 10 K, while the dashed lines represent that of 15 K. 
A lower adsorption energy leads to more efficient surface diffusion, which increase the chance to meet the reaction partner before locked into the ice mantle. 
The characteristic frequency for diffusion and desorption is assumed to be 10$^{12}$ s$^{-1}$. 
The gas density is assumed to be 2 $\times$ 10$^5$ cm$^{-3}$.
See Section \ref{sec_RE_N} for more details. 
}
\label{timescale}
\end{center}
\end{figure}

It should be noted that adsorption energies of other species, particularly those of radicals, are important but uncertain parameters in astrochemical simulations as pointed out in \citet{Wak17}. 
In the low-temperature regime that we consider in this work, diffusion of such species are much less efficient compared to the formation of a monolayer ice (see Fig. \ref{timescale} and discussion in Section \ref{sec_RE_N}), because adsorption energies of major radicals are generally believed to be higher than 1000 K \citep[][and references therein]{Wak17}. 
We thus presume that uncertainties caused by diffusion of high-$E_\text{ads.}$ species would be moderated in the present simulations. 

%%%%%%%%%%
%%%%%%%%%%

%%%%%%%%%%
\begin{deluxetable}{ l c }
\tablecaption{Initial abundances of selected species \label{tab_initab}}
\tablewidth{0pt} 
\tabletypesize{\scriptsize}
\tablehead{
\colhead{Species}   & \colhead{Fractional abundance w.r.t. $n_\mathrm{H}$}  
}
\startdata 
H                                     &     5.0(-5)    \\
H$_2$                             &     5.0(-1)    \\
C$^+$                             &     4.0(-5)       \\
N                                     &    2.5(-5)       \\
O                                     &    1.4(-4)       \\
CO                                  &    4.0(-5)        \\
\enddata
\tablecomments{A(-B) means A $\times$ 10$^{-B}$. 
Elemental abundances are taken from \citet{Aik99}.}
\end{deluxetable}
%%%%%%%%%%

%%%%%%%%%%
\begin{figure*}[tp]
\begin{center}
\includegraphics[width=15cm]{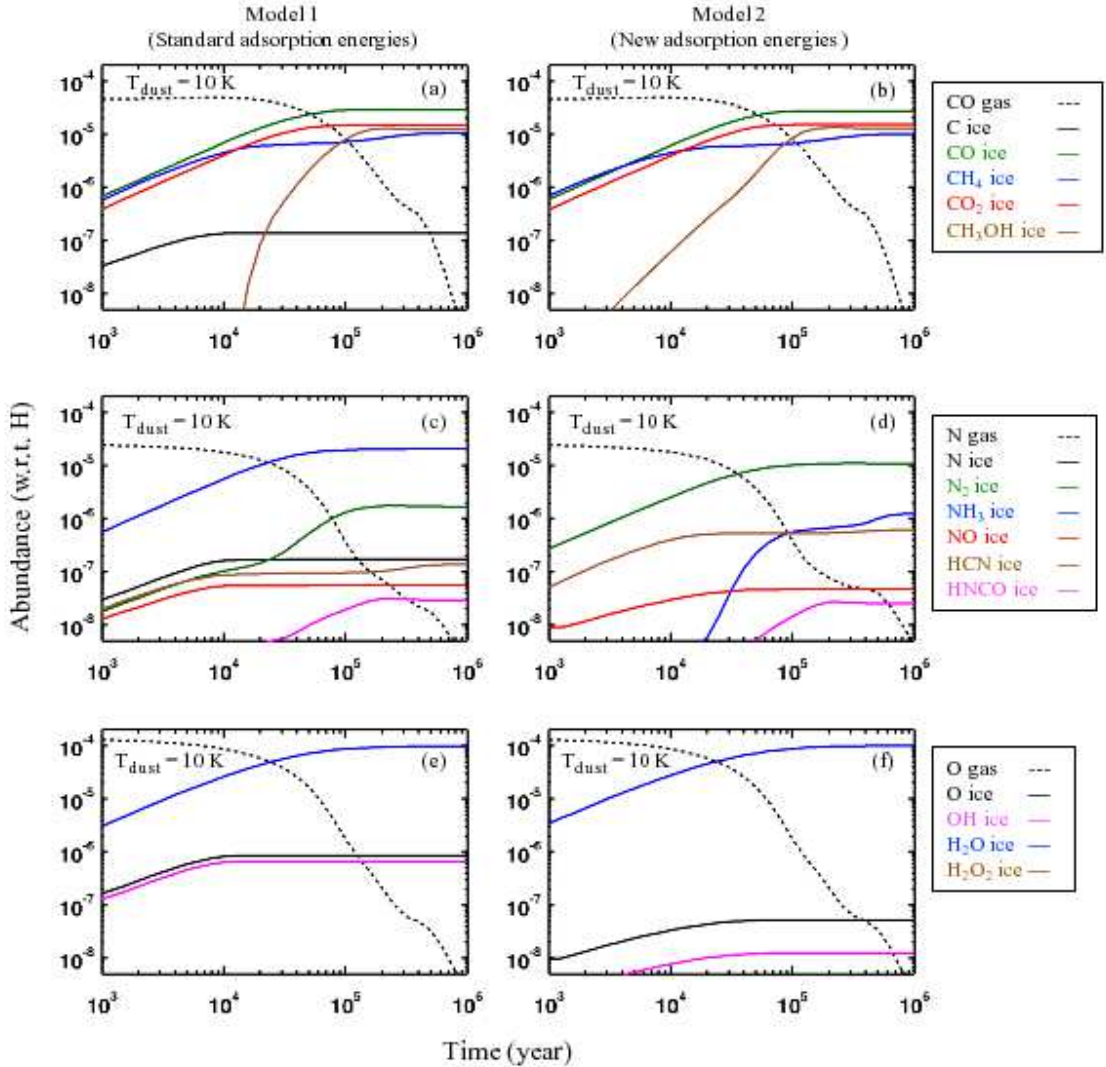}
\caption{
Chemical compositions of a dense molecular cloud ($T_{\mathrm{dust}}$ = 10 K) calculated by the rate equation method using two different sets of adsorption energies; Model 1 (left) and Model 2 (right) (see Section \ref{sec_Method_RE}). 
Time-dependent fractional abundances of important surface species are shown by solid lines in each panel; (a)(b) carbon-bearing species, (c)(d) nitrogen-bearing species, (e)(f) oxygen-bearing species. 
Abundances of major gas-phase species (CO, N, O) are shown by dashed lines. 
}
\label{rate_eq_10K}
\end{center}
\end{figure*}
%%%%%%%%%%
%
%%%%%%%%%%
\begin{figure*}[tp]
\begin{center}
\includegraphics[width=15cm]{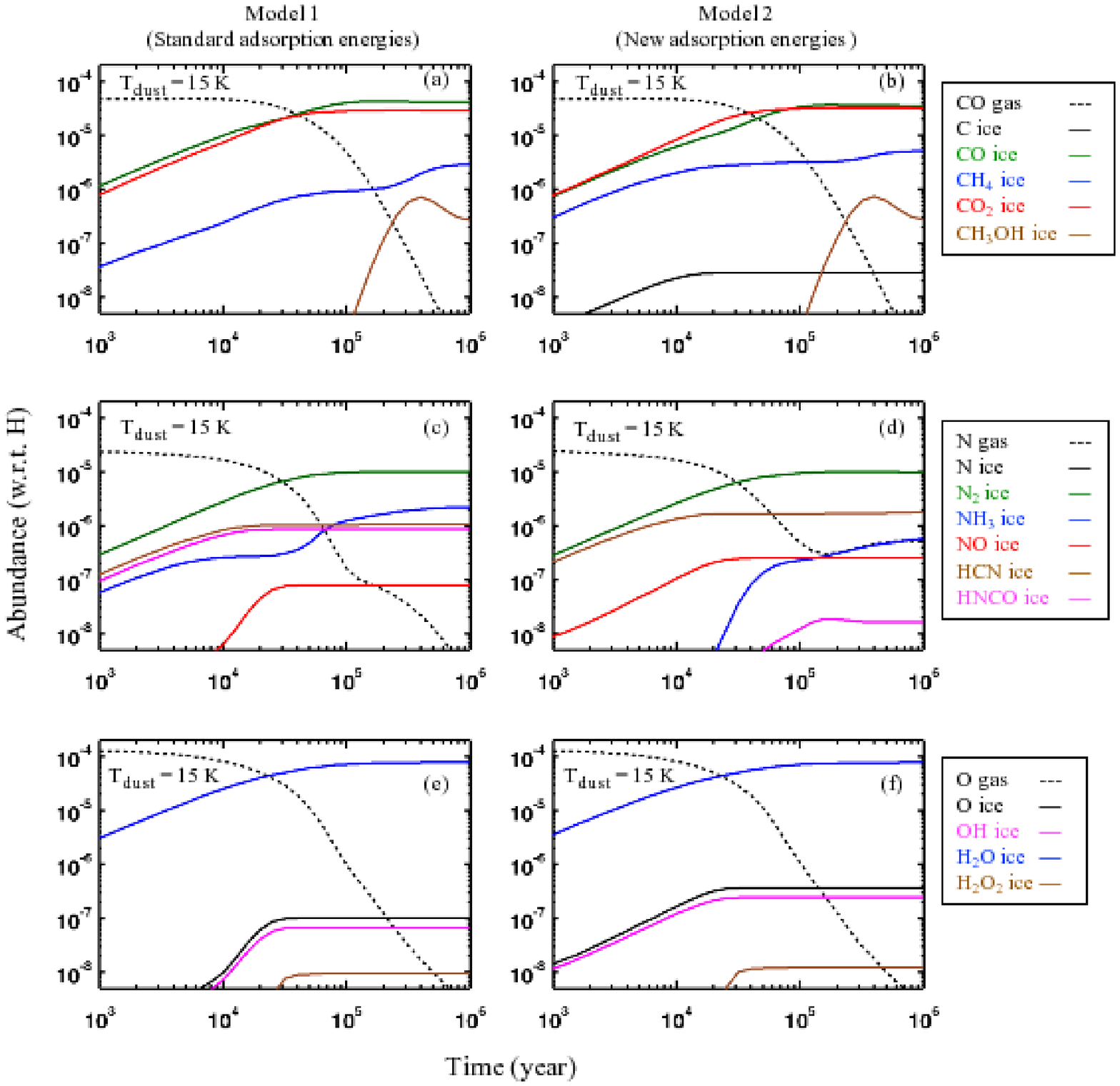}
\caption{Same as Fig. \ref{rate_eq_10K} but for $T_{\mathrm{dust}}$ = 15 K. 
}
\label{rate_eq_15K}
\end{center}
\end{figure*}
%%%%%%%%%%

\subsection{Implication for ice chemistry in dense molecular clouds} \label{sec_Results_RE} 
Figures \ref{rate_eq_10K} and \ref{rate_eq_15K} show the results of gas-ice chemistry simulations at dust temperatures ($T{_\mathrm{dust}}$) of 10 K and 15 K, respectively. 
Fractional abundances with respective to the total hydrogen nuclei density are compared between the model with commonly-used adsorption energies (Model 1) and that with the newly-calculated adsorption energies (Model 2). 
Abundances of selected species in the whole ice mantle extracted at 10$^4$ yr, 10$^5$ yr, and 10$^6$ yr are summarized in Tables \ref{tab_ab_10K}--\ref{tab_ab_15K}. 
The following sections describe effects of the modified adsorption energies on abundances of carbon-, nitrogen-, and oxygen-bearing species in dense clouds.

\subsubsection{Carbon-bearing species} \label{sec_RE_C} 
Most major carbon-bearing species are not significantly affected by the modification of adsorption energies both at 10 K and 15 K (Fig. \ref{rate_eq_10K}ab and Fig. \ref{rate_eq_15K}ab). 
One exception is the atomic carbon ice, whose abundance decrease from Model 1 to Model 2 by about two orders of magnitude at 10 K. 
This would reflect the increased efficiency of the C + N reaction due to the increased surface mobility of atomic nitrogen. 
Note that surface mobility of atomic hydrogen does not change between two models because we use the same adsorption energy in both models. 

On the other hand, at 15 K and in Model 1, the atomic carbon start to diffuse on the grain surface to react with other species, which results in the decreased abundance of surface carbon atoms. 
This pathway is, however, significantly suppressed in Model 2 because the increased adsorption energy of carbon makes the surface carbon atoms almost immobile even at 15 K. 
Therefore, at 15 K, the abundance of the atomic carbon ice is higher in Model 2 than in Model 1. 
The high adsorption energy of carbon implies that its surface diffusion is less efficient until the self-diffusion of water molecules starts at an elevated temperature \citep[$\sim$90 K,][]{Ghe15}. 

The chemisorption of carbon atom on water ice, suggested in our quantum chemistry calculation, implies a possibility of further chemical reactions to produce molecules bearing a C--O bond such as formaldehyde (H$_2$CO) or methanol (CH$_3$OH), while it may hinder the formation of methane (CH$_4$) on water ice via sequential hydrogenation of carbon atoms. 
Although this would have a large impact on the chemical evolution of carbon-bearing species, the present astrochemical model does not include these reaction pathways due to the lack of available chemical network information. 
Further effort to involve those reactions will be an important future work. 

Furthermore, the decreased mobility of atomic carbon would affect the efficiency of organic chemistry that is triggered by the surface diffusion of carbon atoms. 
Several chemical pathways leading to the formation of complex organic molecules through addition reactions of atomic carbon and hydrogen to CO are suggested in the literature \citep[e.g., Figure 12 in][and references therein]{Her09}. 
The present results imply that these pathways, which require the diffusion of atomic carbon (or hydrogenated products of CO), would be less efficient in actual molecular cloud conditions than had been previously thought, due to the decreased mobility of surface carbon atoms. 

%%%%%%%%%%
\begin{deluxetable}{ l c c c c c c }
\tablecaption{Abundances of species in the whole ice mantle for 10 K dust \label{tab_ab_10K}}
\tablewidth{0pt} 
\tabletypesize{\scriptsize} 
\tablehead{
\colhead{}   &  \multicolumn{3}{c}{Model 1\tablenotemark{a}}  &  \multicolumn{3}{|c}{Model 2\tablenotemark{b}}       \\  \cline{2-7} 
\colhead{Species}   & \colhead{10$^4$ yr} & \colhead{10$^5$ yr} & \colhead{10$^6$ yr}  & \colhead{10$^4$ yr} & \colhead{10$^5$ yr} & \colhead{10$^6$ yr}  
}
\startdata 
C                     &   1.4(-7)   &   1.4(-7)   &   1.4(-7)   &   2.6(-9)   &   2.8(-9)   &   2.8(-9)  \\
CO                    &   7.0(-6)   &   2.9(-5)   &   2.9(-5)   &   6.2(-6)   &   2.7(-5)   &   2.7(-5)   \\
CH$_4$                &   4.6(-6)   &   7.2(-6)   &   1.1(-5)   &   4.6(-6)   &   6.8(-6)   &   1.0(-5)   \\
CO$_2$                &   4.1(-6)   &   1.5(-5)   &   1.5(-5)   &   4.0(-6)   &   1.5(-5)   &   1.5(-5)   \\
CH$_3$OH              &   1.2(-10)   &   8.0(-6)   &   1.2(-5)   &   6.3(-8)   &   8.7(-6)   &   1.3(-5)  \\ 
N                     &   1.6(-7)   &   1.7(-7)   &   1.7(-7)   &   5.6(-15)   &   2.9(-14)   &   3.1(-14)   \\
N$_2$                 &   1.0(-7)   &   1.3(-6)   &   1.7(-6)   &   2.6(-6)   &   1.0(-5)   &   1.1(-5)   \\
NH$_3$                &   5.6(-6)   &   2.0(-5)   &   2.1(-5)   &   1.7(-9)   &   5.7(-7)   &   1.3(-6)   \\
NO                    &   5.4(-8)   &   5.6(-8)   &   5.6(-8)   &   3.0(-8)   &   4.7(-8)   &   4.7(-8)   \\
HCN                   &   8.7(-8)   &   9.6(-8)   &   1.4(-7)   &   4.1(-7)   &   5.3(-7)   &   6.2(-7)   \\
HNCO                  &   4.6(-9)   &   2.0(-8)   &   2.9(-8)   &   1.0(-10)   &   1.4(-8)   &   2.5(-8)   \\
O                     &   8.3(-7)   &   8.6(-7)   &   8.6(-7)   &   3.4(-8)   &   5.2(-8)   &   5.2(-8)   \\
O$_2$                 &   8.9(-11)   &   5.7(-10)   &   5.8(-10)   &   4.7(-12)   &   5.9(-10)   &   6.0(-10)   \\
OH                    &   6.5(-7)   &   6.7(-7)   &   6.7(-7)   &   7.9(-9)   &   1.3(-8)   &   1.3(-8)   \\
H$_2$O                &   2.6(-5)   &   8.7(-5)   &   9.7(-5)   &   2.8(-5)   &   8.8(-5)   &   9.9(-5)   \\
H$_2$O$_2$            &   1.6(-10)   &   1.1(-9)   &   1.1(-9)   &   7.8(-12)   &   1.0(-9)   &   1.0(-9)   \\
\enddata
\tablecomments{
A(-B) means A $\times$ 10$^{-B}$. \\
$^a$Standard adsorption energies. 
$^b$New adsorption energies calculated in this work. 
See Section \ref{sec_Method_RE} for details of adsorption energies used in each model. 
}
\end{deluxetable}
%%%%%%%%%%

%%%%%%%%%%
\begin{deluxetable}{ l c c c c c c }
\tablecaption{Abundances of species in the whole ice mantle for 15 K dust \label{tab_ab_15K}}
\tablewidth{0pt} 
%\rotate
\tabletypesize{\scriptsize} 
\tablehead{
\colhead{}   &  \multicolumn{3}{c}{Model 1\tablenotemark{a}}  &  \multicolumn{3}{|c}{Model 2\tablenotemark{b}}       \\  \cline{2-7} 
\colhead{Species}   & \colhead{10$^4$ yr} & \colhead{10$^5$ yr} & \colhead{10$^6$ yr}  & \colhead{10$^4$ yr} & \colhead{10$^5$ yr} & \colhead{10$^6$ yr}  
}
\startdata 
C                     &   3.6(-11)   &   3.7(-11)   &   3.7(-11)   &   2.3(-8)   &   2.9(-8)   &   2.9(-8)  \\
CO                    &   1.0(-5)   &   4.0(-5)   &   4.2(-5)   &   6.2(-6)   &   3.4(-5)   &   3.5(-5)   \\
CH$_4$                &   2.5(-7)   &   9.5(-7)   &   3.0(-6)   &   2.1(-6)   &   3.2(-6)   &   5.2(-6)   \\
CO$_2$                &   7.5(-6)   &   2.8(-5)   &   2.9(-5)   &   8.5(-6)   &   3.2(-5)   &   3.2(-5)   \\
CH$_3$OH              &   2.2(-12)   &   1.9(-9)   &   2.7(-7)   &   5.2(-12)   &   2.3(-9)   &   2.7(-7)   \\
N                     &   6.6(-11)   &   9.0(-11)   &   9.0(-11)   &   5.7(-19)   &   5.3(-18)   &   6.0(-18)   \\
N$_2$                 &   2.8(-6)   &   9.8(-6)   &   1.0(-5)   &   2.7(-6)   &   9.5(-6)   &   9.9(-6)   \\
NH$_3$                &   2.6(-7)   &   1.3(-6)   &   2.2(-6)   &   8.4(-10)   &   2.3(-7)   &   6.0(-7)   \\
NO                    &   6.6(-9)   &   8.0(-8)   &   8.0(-8)   &   1.1(-7)   &   2.5(-7)   &   2.5(-7)   \\
HCN                   &   8.4(-7)   &   1.0(-6)   &   1.1(-6)   &   1.4(-6)   &   1.7(-6)   &   1.8(-6)   \\
HNCO                  &   6.8(-7)   &   8.8(-7)   &   8.8(-7)   &   1.2(-10)   &   1.3(-8)   &   1.6(-8)   \\
O                     &   1.0(-8)   &   1.0(-7)   &   1.0(-7)   &   1.7(-7)   &   3.7(-7)   &   3.7(-7)   \\
O$_2$                 &   4.9(-12)   &   2.8(-9)   &   2.8(-9)   &   4.2(-11)   &   3.7(-9)   &   3.7(-9)   \\
OH                    &   7.4(-9)   &   6.7(-8)   &   6.7(-8)   &   1.2(-7)   &   2.5(-7)   &   2.5(-7)   \\
H$_2$O                &   2.5(-5)   &   7.2(-5)   &   7.8(-5)   &   2.7(-5)   &   7.1(-5)   &   7.7(-5)   \\
H$_2$O$_2$            &   1.7(-11)   &   9.6(-9)   &   9.6(-9)   &   1.4(-10)   &   1.2(-8)   &   1.2(-8)   \\
\enddata
\tablecomments{
A(-B) means A $\times$ 10$^{-B}$. \\
$^a$Standard adsorption energies. 
$^b$New adsorption energies calculated in this work. 
See Section \ref{sec_Method_RE} for details of adsorption energies used in each model. 
}
\end{deluxetable}
%%%%%%%%%%

\subsubsection{Nitrogen-bearing species} \label{sec_RE_N} 
Nitrogen-bearing species are significantly affected by the present modification of adsorption energies. 
At 10 K and in Model 1, the most abundant nitrogen-bearing species is NH$_3$, because hydrogenation dominates grain surface reactions (Fig. \ref{rate_eq_10K}c). 
In Model 2, however, N$_2$ takes the place of the major nitrogen reservoir, because the decreased adsorption energy of atomic nitrogen leads to the effective diffusion, which results in a competition between hydrogenation and \textit{nitrogenation} (Fig. \ref{rate_eq_10K}d). 
In addition, as a consequence of the increased surface reactivity of nitrogen atoms, the abundance of the N ice decreases in Model 2 by nearly seven orders of magnitude compared to Model 1. 
A slight increase of HCN in Model 2 is also seen. 

Figures \ref{timescale} shows timescales of the surface diffusion (scanning of 10$^6$ sites), desorption, and formation of a single ice layer as a function of the adsorption energy of a surface species. 
We here use the inverse of Eq. (4) and (12) in \citet{Cup17} to calculate the plotted timescales. 
The figure indicates that, at 10 K and with $E_\mathrm{ads.}$ = 800 K, both diffusion and desorption are much slower than the formation of a monolayer, thus surface species will hardly have chance to find the reaction parter before being embedded in the mantle phase. 
On the other hand, with $E_\mathrm{ads.}$ = 400 K, the diffusion is much faster than the layer formation and thus surface species have sufficient chance to meet the reaction partner. 
At 15 K, surface species can diffuse rapidly enough to react before the formation of another ice layer even with $E_\mathrm{ads.}$ = 800 K thanks to the elevated temperature. 

An enhancement of surface nitrogenation that are caused by the efficient diffusion of nitrogen atoms provides us an important astrochemical implication. 
So far, the N$_2$ ice has not been detected directly in dense molecular clouds due to the lack of strong infrared bands \citep[e.g.,][]{San01}. 
The suggested formation of N$_2$ as a main reservoir of nitrogen in a dark cloud environment therefore offers an important theoretical implication for the nitrogen budget in dense molecular clouds. 
The universality of the efficient N$_2$ formation, however, should be further investigated for more diverse interstellar conditions. 

\begin{figure}[tbp]
\begin{center}
\includegraphics[width=7.5cm]{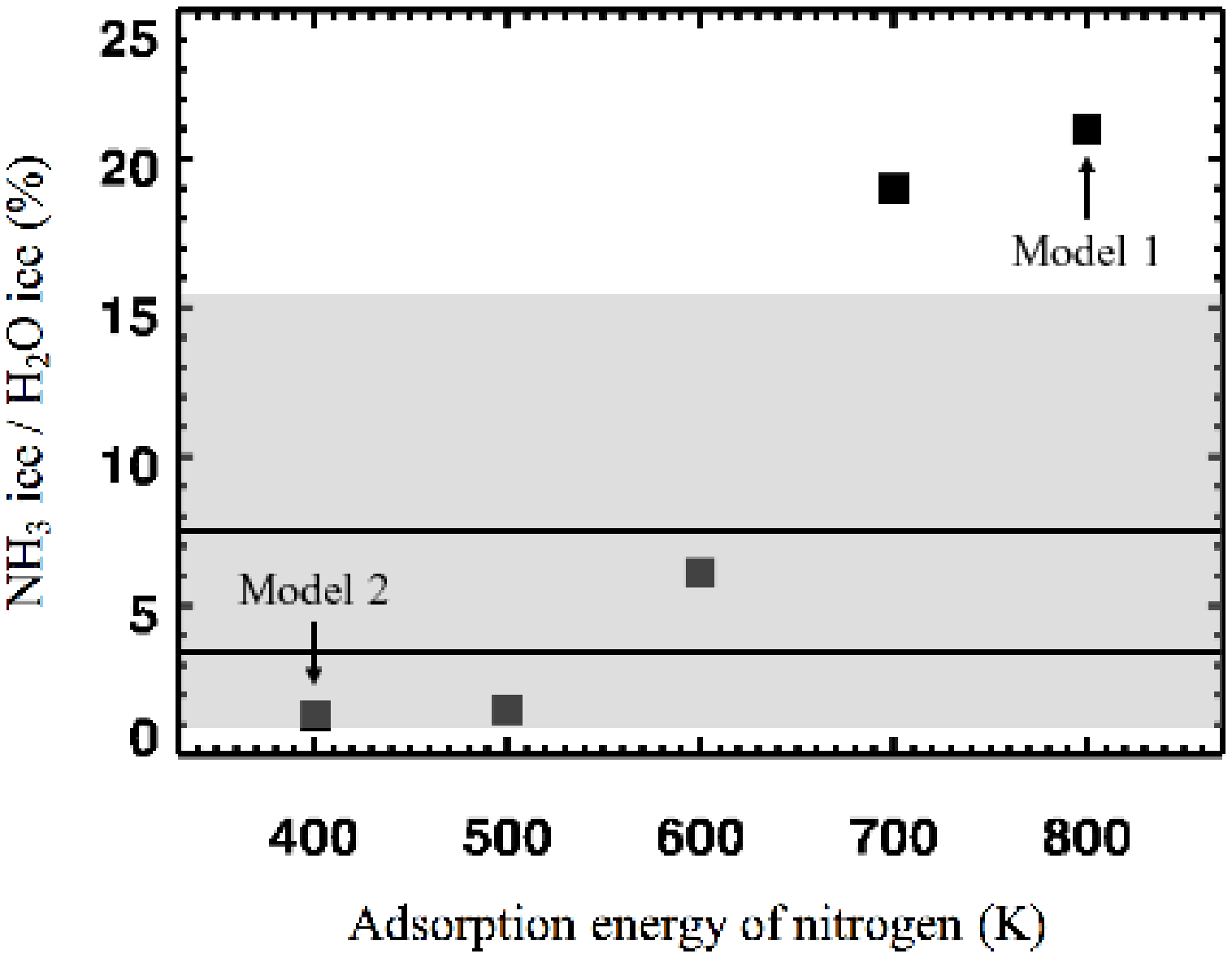}
\caption{Calculated NH$_3$ ice abundances ($T_{\mathrm{dust}}$ = 10 K and time = 10$^6$ year) as a function of the adopted nitrogen adsorption energies (filled squares). 
The shaded area represents the range of the observed NH$_3$ ice abundances towards low-mass and high-mass protostars \citep{Gib01,Dar01, Dar02,Bot10}. 
The two solid lines represent the range of an average and a standard deviation of the NH$_3$ ice abundances for low-mass protostars \citep[5.5 $\pm$ 2.0 $\%$ as reported in][]{Bot10}. 
The adsorption energies of atomic nitrogen used in Model 1 and Model 2 are labeled. 
}
\label{Eb_NH3}
\end{center}
\end{figure}

In general, published gas-ice astrochemical models of dense molecular clouds overestimate the NH$_3$/H$_2$O abundance ratio by a factor of a few \citep[e.g.,][]{Vas13,Cha14,Fur15}, compared to the observationally derived abundance ratio in dense clouds \citep[e.g.,][]{Gib01,Dar01, Dar02,Bot10}. 
The decreased NH$_3$ abundance w.r.t. H$_2$O ice in Model 2 is consistent with the ice observations. 
Figure \ref{Eb_NH3} shows the calculated NH$_3$ abundances as a function of the adopted adsorption energies of atomic nitrogen. 
Observed abundances of the NH$_3$ ice towards low-mass and high-mass protostars are also shown for comparison purpose. 
It is shown that high nitrogen adsorption energies such as in Model 1 overproduces NH$_3$ as compared to the observations, while low adsorption energies as in our quantum chemistry calculations better reproduce the observed NH$_3$ ice abundances. 
Note that the nitrogen adsorption energy of 400 K in Model 2 may underproduce NH$_3$ in some degree compared to the observations since the calculated NH$_3$ abundance is located at the lower end of the observed abundance range. 
This would suggest that the actual adsorption energy of atomic nitrogen might be somewhat higher than the present result as also mentioned in Section \ref{sec_Eb_N}. 

%%%%%%%%%%
%%%%%%%%%%

At 15 K, the dominant nitrogen reservoir is N$_2$ both in Model 1 and 2 because the surface mobility of atomic nitrogen increases at a higher dust temperature (Fig. \ref{rate_eq_15K}cd)). 
At this temperature, the effect of the modified adsorption energies appear as the decreased abundances of NH$_3$ and HNCO, and the increased abundances of HCN and NO. 
A possible explanation on the behaviors of NH$_3$ and HCN is same as in the case of the 10 K simulation described above. 
The slight increase of NO is possibly due to the decreased destruction via NO + C $\to$ OCN, which is caused by the increased adsorption energy of carbon in Model 2. 
Accordingly, the reduced formation of OCN results in the decrease of HNCO, which is formed by OCN + H in our chemical model.

\subsubsection{Oxygen-bearing species} \label{sec_RE_O} 
The most abundant oxygen-bearing species, H$_2$O, is little affected by the modification of adsorption energies both at 10 K and 15 K (Fig. \ref{rate_eq_10K}ef and Fig. \ref{rate_eq_15K}ef). 
Behaviors of relatively minor solid species, O and OH, are rather complicated; their abundances decrease from Model 1 to Model 2 at 10 K, while they increase from Model 1 to Model 2 at 15 K. 
Because the oxygen adsorption energies used in Model 1 and Model 2 are close, abundance differences seen in those oxygen-bearing species are likely due to the indirect effect caused by the modifications of carbon and nitrogen adsorption energies. 
The decrease of O and OH from Model 1 to Model 2 at 10 K may be due to the enhancement of N + O reaction in Model 2, while the increase of O and OH from Model 1 to Model 2 at 15 K may reflect the suppression of the O production by the diffusive surface reaction of carbon- and oxygen-bearing species. 
Note that the O$_2$ ice is not shown in the figures because its abundance is always lower than 10$^{-8}$ in our calculation (Tables \ref{tab_ab_10K}--\ref{tab_ab_15K}).

\section{Conclusions} \label{sec_concl}
We carried out the quantum chemistry calculations to estimate adsorption energies of atomic carbon, nitrogen, and oxygen on the low-temperature amorphous water ice surface. 
In addition, we investigate the effect of the newly calculated adsorption energies on chemical compositions of dense molecular clouds with the aid of the gas-ice astrochemistry simulation. 
We obtain the following conclusions in this work. 

\begin{enumerate}
\item
A new computational model which is able to merge quantum chemical and statistical contributions to binding energies of atoms and molecules in interstellar chemistry is proposed. 
An idea of our computational model is that an atom or a molecule adsorbs to locally the most stable site on ASW surface during the very-long period of time. 
This is achieved by considering variation of adsorption sites on ASW by sampling structures of water clusters. 
Therefore, it is advantageous that small computational efforts are only required, and this is because quantum chemical calculations are available in our computational model. 
Although our computational results might not be converged from statistical viewpoint since we only sampled a small number of water clusters for demonstration, we conclude that our model is quite useful to theoretically estimate binding energies in interstellar chemistry. 

\item
The calculated adsorption energies of atoms on ASW are 14100 $\pm$ 420 K for carbon, 400 $\pm$ 30 K for nitrogen, 1440 $\pm$ 160 K for oxygen. 
The estimated binding energy of oxygen agrees well with the experimental numbers. 
An N atom takes purely physisorption, and therefore, the binding energy of an N atom is amazingly small. 
On the other hand, a C atom apparently shows chemisorption to form a chemical bond between an O atom in water molecule. 
An O atom has a dual character of both physisorption and chemisorption. 
In consequence, the binding energies are in the order, C $\gg$ O $>$ N. 

\item
The high adsorption energy of carbon suggests that its surface diffusion is less efficient until water molecules start to diffuse at high temperature. 
The decreased mobility of carbon would suppress the previously suggested formation pathways of complex organic molecules which is triggered by the surface diffusion of atomic carbon. 
On the other hand, the chemisorption of the C atom also suggests a possibility of further chemical reactions to produce molecules bearing a C--O bond such as formaldehyde, methanol, and so on. 
This would be of a large impact to the chemical evolution of carbon species, and we are going to extend our model to involve such reactions in astrochemical simulations as a future work. 
This would be of a large impact to the chemical evolution of carbon-bearing species in dense molecular clouds. 

\item
The low adsorption energy of nitrogen implies that atomic nitrogen can efficiently diffuse on the surface even at 10 K. 
This significantly alters chemical compositions of nitrogen-bearing molecules in dense molecular clouds. 
The most notable effect is that the N$_2$ is formed as a main reservoir of nitrogen in stead of NH$_3$ at low temperatures, because surface nitrogenation competes with hydrogenation. 

\item
Major oxygen-bearing surface species are little affected by the application of the new adsorption energies. 
\end{enumerate}

Future work will need to investigate adsorption energies of further atoms and molecules of astrochemical interest.

\acknowledgments
This work is supported by a Grant-in-Aid from the Japan Society for the Promotion of Science (15K17612), and Building of Consortia for the Development of Human Resources in Science and Technology, MEXT, Japan. 
The authors are grateful to Prof. Yuri Aikawa and Prof. Naoki Watanabe for fruitful discussions and helpful suggestions. 
Finally, we would like to thank an anonymous referee for careful reading and helpful comments. 
\software{Amber 14 \citep{amber14}, Gaussian 09 \citep{gaussian}}

\clearpage

\appendix
\section{Adsorption energies of C, N, and O atoms estimated by different calculation methods} 
Table \ref{tab_A1} summarizes adsorption energies of carbon, nitrogen, and oxygen atoms to one water molecule estimated by three different calculation methods. 
See discussion in Section \ref{sec_Eb_N}.

%%%%%%%%%%
\begin{deluxetable}{ l c c c c }
\tablecaption{
Adsorption energies of C, N, and O atoms to one water molecule, estimated by using DFT ($\omega$B97XD/6-311+G(d, p)), MP2/aug-cc-pVQZ, and CCSD(T)/aug-cc-pVQZ levels of theory \label{tab_A1}
}
\tabletypesize{\small} 
\tablehead{
\colhead{}               &  \multicolumn{3}{c}{Adsorption energy (K)}       \\  \cline{2-4} 
\colhead{Species}  & \colhead{$\omega$B97XD} & \colhead{MP2} & \colhead{CCSD(T)\tablenotemark{a}} 
}
\startdata 
C--OH$_2$       & 5708          & 4289  & 4314 \\
C--HOH            & 497            & 594    & 605 \\
N--OH$_2$      & Not bound  & 105    & 124 \\
N--HOH            & 131            & 163    & 183 \\
O--OH$_2$      & 926             & 473    & 528 \\
O--HOH           & 823             & 842    & 879 \\
\enddata
\tablecomments{
Species "X−OH$_2$" and "X−HOH" (X = C, N, and O) denote whether X atom binds to O atom or H atom of the water molecule.
$^a$CCSD(T) calculations were carried out based on the MP2-optimized geometry. 
}
\end{deluxetable}
%%%%%%%%%%

%\clearpage

%\bibliographystyle{aasjournal}
%\bibliography{references}

\begin{thebibliography}{}
\expandafter\ifx\csname natexlab\endcsname\relax\def\natexlab#1{#1}\fi
\providecommand{\url}[1]{\href{#1}{#1}}

\bibitem[{Ahmed {et~al.}(1983)Ahmed, McKee, \& Shevlin}]{Ahm83}
Ahmed, S.~N., McKee, M.~L., \& Shevlin, P.~B. 1983, Journal of the American
  Chemical Society, 105, 3942

\bibitem[{{Aikawa} \& {Herbst}(1999)}]{Aik99}
{Aikawa}, Y., \& {Herbst}, E. 1999, \apj, 526, 314

\bibitem[{Al-Halabi {et~al.}(2002)Al-Halabi, Kleyn, Van~Dishoeck, \&
  Kroes}]{AlH02}
Al-Halabi, A., Kleyn, A., Van~Dishoeck, E., \& Kroes, G. 2002, The Journal of
  Physical Chemistry B, 106, 6515

\bibitem[{{Al-Halabi} \& {van Dishoeck}(2007)}]{AlH07}
{Al-Halabi}, A., \& {van Dishoeck}, E.~F. 2007, \mnras, 382, 1648

\bibitem[{Amiaud {et~al.}(2007)Amiaud, Dulieu, Fillion, Momeni, \&
  Lemaire}]{Ami07}
Amiaud, L., Dulieu, F., Fillion, J.-H., Momeni, A., \& Lemaire, J. 2007, The
  Journal of chemical physics, 127, 144709

\bibitem[{Amiaud {et~al.}(2006)Amiaud, Fillion, Baouche, Dulieu, Momeni, \&
  Lemaire}]{Ami06}
Amiaud, L., Fillion, J., Baouche, S., {et~al.} 2006, The Journal of chemical
  physics, 124, 094702

\bibitem[{{\'A}sgeirsson {et~al.}(2017){\'A}sgeirsson, J{\'o}nsson, \&
  Wikfeldt}]{Asg17}
{\'A}sgeirsson, V., J{\'o}nsson, H., \& Wikfeldt, K. 2017, The Journal of
  Physical Chemistry C, 121, 1648

\bibitem[{{Bottinelli} {et~al.}(2010){Bottinelli}, {Boogert}, {Bouwman},
  {Beckwith}, {van Dishoeck}, {{\"O}berg}, {Pontoppidan}, {Linnartz}, {Blake},
  {Evans}, \& {Lahuis}}]{Bot10}
{Bottinelli}, S., {Boogert}, A.~C.~A., {Bouwman}, J., {et~al.} 2010, \apj, 718,
  1100

\bibitem[{Buch \& Czerminski(1991)}]{Buc91}
Buch, V., \& Czerminski, R. 1991, The Journal of chemical physics, 95, 6026

\bibitem[{Burke \& Brown(2010)}]{Bur10}
Burke, D.~J., \& Brown, W.~A. 2010, Physical Chemistry Chemical Physics, 12,
  5947

\bibitem[{Chai \& Head-Gordon(2008)}]{chai2008}
Chai, J.-D., \& Head-Gordon, M. 2008, Physical Chemistry Chemical Physics, 10,
  6615

\bibitem[{{Chang} \& {Herbst}(2014)}]{Cha14}
{Chang}, Q., \& {Herbst}, E. 2014, \apj, 787, 135

\bibitem[{{Cuppen} {et~al.}(2017){Cuppen}, {Walsh}, {Lamberts}, {Semenov},
  {Garrod}, {Penteado}, \& {Ioppolo}}]{Cup17}
{Cuppen}, H.~M., {Walsh}, C., {Lamberts}, T., {et~al.} 2017, \ssr, 212, 1

\bibitem[{{Dalgarno}(2006)}]{Dal06}
{Dalgarno}, A. 2006, Proceedings of the National Academy of Science, 103, 12269

\bibitem[{{Dartois} \& {d'Hendecourt}(2001)}]{Dar01}
{Dartois}, E., \& {d'Hendecourt}, L. 2001, \aap, 365, 144

\bibitem[{{Dartois} {et~al.}(2002){Dartois}, {d'Hendecourt}, {Thi},
  {Pontoppidan}, \& {van Dishoeck}}]{Dar02}
{Dartois}, E., {d'Hendecourt}, L., {Thi}, W., {Pontoppidan}, K.~M., \& {van
  Dishoeck}, E.~F. 2002, \aap, 394, 1057

\bibitem[{{Dupuy} {et~al.}(2016){Dupuy}, {Lewis}, \& {Stancil}}]{Dup16}
{Dupuy}, J.~L., {Lewis}, S.~P., \& {Stancil}, P.~C. 2016, \apj, 831, 54

\bibitem[{Fillion {et~al.}(2009)Fillion, Amiaud, Congiu, Dulieu, Momeni, \&
  Lemaire}]{Fil09}
Fillion, J.-H., Amiaud, L., Congiu, E., {et~al.} 2009, Physical Chemistry
  Chemical Physics, 11, 4396

\bibitem[{{Frisch}(2015)}]{gaussian}
{Frisch}, M.~J. 2015, Gaussian 09, http://gaussian.com/ 

\bibitem[{{Furuya} {et~al.}(2015){Furuya}, {Aikawa}, {Hincelin}, {Hassel},
  {Bergin}, {Vasyunin}, \& {Herbst}}]{Fur15}
{Furuya}, K., {Aikawa}, Y., {Hincelin}, U., {et~al.} 2015, \aap, 584, A124

\bibitem[{{Furuya} {et~al.}(2017){Furuya}, {Drozdovskaya}, {Visser}, {van
  Dishoeck}, {Walsh}, {Harsono}, {Hincelin}, \& {Taquet}}]{Fur17}
{Furuya}, K., {Drozdovskaya}, M.~N., {Visser}, R., {et~al.} 2017, \aap, 599,
  A40

\bibitem[{{Furuya} {et~al.}(2016){Furuya}, {van Dishoeck}, \& {Aikawa}}]{Fur16}
{Furuya}, K., {van Dishoeck}, E.~F., \& {Aikawa}, Y. 2016, \aap, 586, A127

\bibitem[{{Garrod}(2008)}]{Gar08a}
{Garrod}, R.~T. 2008, \aap, 491, 239

\bibitem[{{Garrod} \& {Herbst}(2006)}]{Gar06}
{Garrod}, R.~T., \& {Herbst}, E. 2006, \aap, 457, 927

\bibitem[{Ghesqui{\`e}re {et~al.}(2015)Ghesqui{\`e}re, Mineva, Talbi,
  Theul{\'e}, Noble, \& Chiavassa}]{Ghe15}
Ghesqui{\`e}re, P., Mineva, T., Talbi, D., {et~al.} 2015, Physical Chemistry
  Chemical Physics, 17, 11455

\bibitem[{{Gibb} {et~al.}(2001){Gibb}, {Whittet}, \& {Chiar}}]{Gib01}
{Gibb}, E.~L., {Whittet}, D.~C.~B., \& {Chiar}, J.~E. 2001, \apj, 558, 702

\bibitem[{{Hama} {et~al.}(2012){Hama}, {Kuwahata}, {Watanabe}, {Kouchi},
  {Kimura}, {Chigai}, \& {Pirronello}}]{Ham12}
{Hama}, T., {Kuwahata}, K., {Watanabe}, N., {et~al.} 2012, \apj, 757, 185

\bibitem[{Hama \& Watanabe(2013)}]{Ham13}
Hama, T., \& Watanabe, N. 2013, Chemical reviews, 113, 8783

\bibitem[{{Hama} \& {Watanabe}(2013)}]{HW13}
{Hama}, T., \& {Watanabe}, N. 2013, Chemical Reviews, 113, 8783

\bibitem[{{Hasegawa} \& {Herbst}(1993)}]{Has93}
{Hasegawa}, T.~I., \& {Herbst}, E. 1993, \mnras, 263, 589

\bibitem[{{Hasegawa} {et~al.}(1992){Hasegawa}, {Herbst}, \& {Leung}}]{Has92}
{Hasegawa}, T.~I., {Herbst}, E., \& {Leung}, C.~M. 1992, \apjs, 82, 167

\bibitem[{{He} {et~al.}(2015){He}, {Shi}, {Hopkins}, {Vidali}, \&
  {Kaufman}}]{He15}
{He}, J., {Shi}, J., {Hopkins}, T., {Vidali}, G., \& {Kaufman}, M.~J. 2015,
  \apj, 801, 120

\bibitem[{{Herbst} \& {van Dishoeck}(2009)}]{Her09}
{Herbst}, E., \& {van Dishoeck}, E.~F. 2009, \araa, 47, 427

\bibitem[{Hickson {et~al.}(2016)Hickson, Loison, Nu{\~n}ez-Reyes, \&
  M{\'e}reau}]{Hic16}
Hickson, K.~M., Loison, J.-C., Nu{\~n}ez-Reyes, D., \& M{\'e}reau, R. 2016, The
  journal of physical chemistry letters, 7, 3641

\bibitem[{Hornek{\ae}r {et~al.}(2003)Hornek{\ae}r, Baurichter, Petrunin, Field,
  \& Luntz}]{Hor03}
Hornek{\ae}r, L., Baurichter, A., Petrunin, V., Field, D., \& Luntz, A. 2003,
  Science, 302, 1943

\bibitem[{Hornek{\ae}r {et~al.}(2005)Hornek{\ae}r, Baurichter, Petrunin, Luntz,
  Kay, \& Al-Halabi}]{Hor05}
Hornek{\ae}r, L., Baurichter, A., Petrunin, V., {et~al.} 2005, The Journal of
  chemical physics, 122, 124701

\bibitem[{{Karssemeijer} \& {Cuppen}(2014)}]{Kar14}
{Karssemeijer}, L.~J., \& {Cuppen}, H.~M. 2014, \aap, 569, A107

\bibitem[{{Kimber} {et~al.}(2014){Kimber}, {Ennis}, \& {Price}}]{Kim14}
{Kimber}, H.~J., {Ennis}, C.~P., \& {Price}, S.~D. 2014, Faraday Discussions,
  168, 167

\bibitem[{Kimmel {et~al.}(2001)Kimmel, Stevenson, Dohnalek, Smith, \&
  Kay}]{Kim01}
Kimmel, G.~A., Stevenson, K.~P., Dohnalek, Z., Smith, R.~S., \& Kay, B.~D.
  2001, The Journal of Chemical Physics, 114, 5284

\bibitem[{{Kollman}(2014)}]{amber14}
{Kollman}, P.~A. 2014, AMBER 14, http://ambermd.org/

\bibitem[{Krasnokutski \& Huisken(2014)}]{Kra14}
Krasnokutski, S.~A., \& Huisken, F. 2014, The Journal of chemical physics, 141,
  214306

\bibitem[{Kuwahata {et~al.}(2015)Kuwahata, Hama, Kouchi, \& Watanabe}]{Kuw15}
Kuwahata, K., Hama, T., Kouchi, A., \& Watanabe, N. 2015, Physical review
  letters, 115, 133201

\bibitem[{Lee \& Meuwly(2014)}]{MLee14}
Lee, M.~W., \& Meuwly, M. 2014, Faraday discussions, 168, 205

\bibitem[{{Manic{\`o}} {et~al.}(2001){Manic{\`o}}, {Ragun{\`i}}, {Pirronello},
  {Roser}, \& {Vidali}}]{Man01}
{Manic{\`o}}, G., {Ragun{\`i}}, G., {Pirronello}, V., {Roser}, J.~E., \&
  {Vidali}, G. 2001, \apjl, 548, L253

\bibitem[{{Matar} {et~al.}(2008){Matar}, {Congiu}, {Dulieu}, {Momeni}, \&
  {Lemaire}}]{Mat08}
{Matar}, E., {Congiu}, E., {Dulieu}, F., {Momeni}, A., \& {Lemaire}, J.~L.
  2008, \aap, 492, L17

\bibitem[{Medve{\v{d}} \& {\v{C}}ern{\`y}(2011)}]{Med11}
Medve{\v{d}}, I., \& {\v{C}}ern{\`y}, R. 2011, Microporous and Mesoporous
  Materials, 142, 405

\bibitem[{Minissale {et~al.}(2016)Minissale, Congiu, \& Dulieu}]{Min16}
Minissale, M., Congiu, E., \& Dulieu, F. 2016, Astronomy \& Astrophysics, 585,
  A146

\bibitem[{Ortman {et~al.}(1990)Ortman, Hauge, Margrave, \& Kafafi}]{Ort90}
Ortman, B.~J., Hauge, R.~H., Margrave, J.~L., \& Kafafi, Z.~H. 1990, Journal of
  Physical Chemistry, 94, 7973

\bibitem[{Ozkan \& Dede(2012)}]{Ozk12}
Ozkan, I., \& Dede, Y. 2012, International Journal of Quantum Chemistry, 112,
  1165

\bibitem[{{Penteado} {et~al.}(2017){Penteado}, {Walsh}, \& {Cuppen}}]{Pen17}
{Penteado}, E.~M., {Walsh}, C., \& {Cuppen}, H.~M. 2017, \apj, 844, 71

\bibitem[{{Perets} {et~al.}(2005){Perets}, {Biham}, {Manic{\'o}}, {Pirronello},
  {Roser}, {Swords}, \& {Vidali}}]{Per05}
{Perets}, H.~B., {Biham}, O., {Manic{\'o}}, G., {et~al.} 2005, \apj, 627, 850

\bibitem[{{Pirronello} {et~al.}(2004){Pirronello}, {Manic{\'o}}, {Roser}, \&
  {Vidali}}]{Pir04}
{Pirronello}, V., {Manic{\'o}}, G., {Roser}, J., \& {Vidali}, G. 2004, in
  Astronomical Society of the Pacific Conference Series, Vol. 309, Astrophysics
  of Dust, ed. A.~N. {Witt}, G.~C. {Clayton}, \& B.~T. {Draine}, 529

\bibitem[{{Sandford} {et~al.}(2001){Sandford}, {Bernstein}, {Allamandola},
  {Goorvitch}, \& {Teixeira}}]{San01}
{Sandford}, S.~A., {Bernstein}, M.~P., {Allamandola}, L.~J., {Goorvitch}, D.,
  \& {Teixeira}, T.~C.~V.~S. 2001, \apj, 548, 836

\bibitem[{Schreiner \& Reisenauer(2006)}]{Sch06}
Schreiner, P.~R., \& Reisenauer, H.~P. 2006, ChemPhysChem, 7, 880

\bibitem[{Senevirathne {et~al.}(2017)Senevirathne, Andersson, Dulieu, \&
  Nyman}]{Sen17}
Senevirathne, B., Andersson, S., Dulieu, F., \& Nyman, G. 2017, Molecular
  Astrophysics, 6, 59

\bibitem[{Sladek {et~al.}(1974)Sladek, Gilliland, \& Baddour}]{Sla74}
Sladek, K.~J., Gilliland, E.~R., \& Baddour, R.~F. 1974, Industrial \&
  Engineering Chemistry Fundamentals, 13, 100

\bibitem[{{Tielens} \& {Hagen}(1982)}]{Tie82}
{Tielens}, A.~G.~G.~M., \& {Hagen}, W. 1982, \aap, 114, 245

\bibitem[{{Vasyunin} \& {Herbst}(2013)}]{Vas13}
{Vasyunin}, A.~I., \& {Herbst}, E. 2013, \apj, 762, 86

\bibitem[{{Veeraghattam} {et~al.}(2014){Veeraghattam}, {Manrodt}, {Lewis}, \&
  {Stancil}}]{Vee14}
{Veeraghattam}, V.~K., {Manrodt}, K., {Lewis}, S.~P., \& {Stancil}, P.~C. 2014,
  \apj, 790, 4

\bibitem[{Vidali {et~al.}(2006)Vidali, Roser, Ling, Congiu, Manic{\'o}, \&
  Pirronello}]{Vid06}
Vidali, G., Roser, J.~E., Ling, L., {et~al.} 2006, Faraday discussions, 133,
  125

\bibitem[{{Wakelam} {et~al.}(2017){Wakelam}, {Loison}, {Mereau}, \&
  {Ruaud}}]{Wak17}
{Wakelam}, V., {Loison}, J.-C., {Mereau}, R., \& {Ruaud}, M. 2017, Molecular
  Astrophysics, 6, 22

\bibitem[{{Ward} {et~al.}(2012){Ward}, {Hogg}, \& {Price}}]{War12}
{Ward}, M.~D., {Hogg}, I.~A., \& {Price}, S.~D. 2012, \mnras, 425, 1264

\bibitem[{{Watanabe} {et~al.}(2010){Watanabe}, {Kimura}, {Kouchi}, {Chigai},
  {Hama}, \& {Pirronello}}]{Wat10}
{Watanabe}, N., {Kimura}, Y., {Kouchi}, A., {et~al.} 2010, \apjl, 714, L233

\end{thebibliography}

\end{document}